\documentstyle[prl,aps,epsf,twocolumn]{revtex}

\tighten

\begin{document}
\draft
\twocolumn[\hsize\textwidth\columnwidth\hsize\csname 
@twocolumnfalse\endcsname

\title{\bf  Cooperative motion in Lennard-Jones binary
mixtures below the glass transition}

\author{Normand Mousseau}

\address{ Department of Physics and Astronomy and CMSS, Ohio University, Athens, OH 45701, 
USA\protect\cite{mousadd}\\
and Computational Physics, Faculty of Applied Physics, TU Delft, 
Lorentzweg 1, 2628 CJ Delft,
The Netherlands
}

\date{\today}

\maketitle



\begin{abstract}

Using the activation-relaxation technique (ART), we study the nature 
of relaxation events in a binary Lennard-Jones system above and below 
the glass transition temperature ($T_g$).  ART generates 
trajectories with almost identical efficiency at all temperature, 
thus avoiding the exponential slowing down below $T_g$ and providing 
extensive sampling everywhere.  Comparing these runs, we find that 
the number of atoms involved in an event decreases strongly with 
temperature.  In particular, while in the supercooled liquid 
activated events are collective, involving on average thirty atoms or 
more, events below T$_g$ involve mostly single atoms and produce 
minimal disturbance of the local environment. These results confirm 
the interpretation and the generality of recent NMR results by Tang 
{\it et al} (Nature {\bf 402}, 160 (1999)).

\end{abstract}

\vskip2pc
]

\vspace*{-0.8cm} \narrowtext

Atomic motion in solids is largely determined by the nature of the 
local network. While atomic diffusion in crystals is constrained by 
symmetry and can be described in terms of well-defined displacements 
leaving the overall structure of the network unaffected, diffusion in 
disordered materials offers a much more complex picture. These 
materials present a wide range of local environments and diffusion  
can take place in principle  through an equally large number of 
mechanisms.

This has made understanding the nature of diffusion and relaxation
mechanisms in compact glasses, such as metallic and Lennard-Jones
glasses, a difficult task.  Nevertheless, significant progress
regarding the details of the dynamics in these dense glasses has been
achieved recently.  On the theoretical side, simulations have been
used extensively to investigate this
problem.~\cite{lewis91,wahnstrom91,heuer,delaye94,kob95a} Studies on
supercooled model binary glasses have established clearly that as a
liquid becomes supercooled a change in the dynamics takes place and
diffusion starts to proceed by jumps.~\cite{lewis91,wahnstrom91} More
recent work has provided further characterization of these mechanisms,
showing that in the supercooled regime moves are collective
\cite{heuer,delaye94} and occur in highly correlated
sequences.~\cite{donati98}

Experimental measurements are also challenging; diffusion takes place
on a long time scale and the data represent only an average over a
distribution of barriers and pre-factors.  This make it difficult to
identify directly specific mechanisms.  A search for diffusion
mechanisms in multinary metallic glasses, using standard slicing
techniques, has led to conflicting
results.~\cite{grandjean97,klugkist98} More direct probes to identify
local changes around atoms, such as NMR, have also been used
recently.~\cite{heuer95,tracht98,tang98,tang99,israelof} In
particular, the work of Tang {\it et al.}~\cite{tang99} implies that
there is a qualitative change in the diffusion mechanism of Be atoms
as the Zr-Ti-Cu-Ni-Be samples are brought below the glass transition
temperature ($T_g$): from mostly collective, jumps become localized
and clearly atomistic.  This last result is of great interest because
it demonstrates a qualitative distinction between the supercooled
regime and the dynamics below $T_g$.

A numerical reproduction of this phenomenon, necessary to establish 
the validity of the explanation and its generality, is difficult to 
achieve using standard techniques: in the low temperature regime, the 
time scale covered by molecular dynamics is insufficient to ensure a 
satisfactory exploration of the space of configurations.  The 
activation-relaxation technique (ART)~\cite{barkema96} offers a way 
to go beyond these limitations and to sample the phase space of 
disordered systems even at low temperatures.

We show here that ART can generate trajectories in Lennard-Jones 
glasses without suffering from exponential slowing down below $T_g$.  
Comparing events above and below $T_g$, we also find that the number 
of atoms involved in relaxation and diffusion decreases significantly 
with temperature, going from many tens to one or two at the lowest 
$T$ studied here.

ART by-passes the description of the phonon vibrations to concentrate
on activated mechanisms: it looks directly for paths connecting minima
in a high-dimensional energy landscape.  Starting from a local
minimum, the whole configuration is first pushed away from it, until a
negative eigenvalue appear, and then directed to a nearby saddle point
-- the activation.  The configuration is then brought to a new
minimum, providing a complete event with initial, saddle and final
configurations.  The new move is then accepted or rejected with
Bolztmann probability ($\exp( \Delta E/k_BT$, where $\Delta E$ is the
energy difference between final and initial configuration and $T$ the 
simulation temperature.)  A more
detailed description of the original algorithm can be found in Ref. 
~\cite{artb}.

In order to ensure a better control of the trajectory in this dense
material, we use here a modified version of the algorithm.  In the
activation stage, the configuration is now pushed against the force
corresponding to the to the lowest (negative) eigenvalue of the
hessian matrix, the second derivative of the total configurational
energy.  Since an exact diagonalization of the $3N \times 3N$ matrix
is computationally too demanding for the 1000-atom simulation
presented here, a Lanczos algorithm is used to project out
eigenvectors corresponding to the lowest eigenvalues
only.~\cite{drabold} In spite of the efficiency of the Lanczos
algorithm, this approach remains numerically more intensive than the
standard ART. However, ART nouveau ensures a direct convergence to the
saddle point and provides a better control on the trajectory, which is
particularly useful in dense systems such as metallic or Lennard-Jones
glasses.

This algorithm is applied to a 1000-atom Lennard-Jones binary
$A_{80}B_{20}$ mixture using the parameters introduced in
Ref.~\cite{kob95a} but with {\it shifted energy and forces} to ensure
continuity of the energy and the first derivative at the
cut-off.~\cite{lj} Energy has units of $\epsilon_{AA}$, temperature of
$\epsilon_{AA}/k_B$, length of $\sigma_{AA}$ and time of
$(m\sigma_{AA}/48 \epsilon_{AA})$.

The simulation procedure is as follows.  We start with a
randomly-packed configuration and first relax it at constant volume
and $T=0.50$, a temperature slightly above the glass transition, until
thermalization, i.e., until the configurational energy reaches a plateau. 
This takes place in about 5000 ART-events.  The last configuration of
the initialization run is then used as a starting point for further
runs at $T=$ 0.25, 0.50 and 1.00, i.e., well below, slightly above and
well above $T_g$, respectively.  $T_g$, here, is defined as in Ref. 
\cite{sastry98}, by a sharp low-temperature break in the
inherent-structure energy curve.~\cite{inprep}

The overall acceptation ratio for these runs is about 20 ~\%.  This
includes exchange events, accounting for about 6~ \% at all
temperatures, where two or more atoms switch position, leaving the
final configuration structurally indistinguishable from the original. 
Although physically relevant in the study of self-diffusion, these
events do not contribute to the relaxation of the lattice {\it per se}
and they excluded from the analysis below.  Accepted events are
therefore only those resulting in a final configuration structurally
{\it different} for the initial one.  The acceptation rate for these
events is about 20 \% at $T=1.00$, 10 \% at $T=0.50$ and 4-5~\% at
$T=0.25$.

\begin{figure}
\vspace*{-0.2cm}
     \epsfxsize=7.7cm
\epsfbox{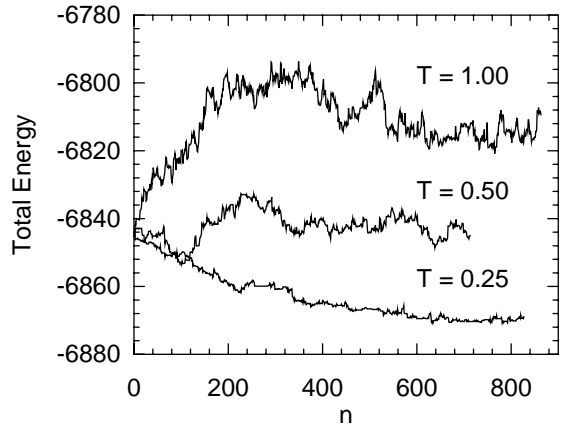}

\vspace{0.3cm} 
\caption{ Configurational energy sequence of the
accepted events, i.e., after removing atomic exchanges.  The
thermalization for T=1.00 and T=0.25 takes about 300 and 500 events
respectively.}
\label{fig:energy}
\end{figure}

Figure \ref{fig:energy} shows the energy sequence of the accepted
events at the three temperatures considered here.  As the model is
initially prepared at T=0.50, this sequence is already thermalized. 
The thermalization at $T=1.00$ takes about 300 events while that at
$T=0.25$ is longer, about 500 events.  It is not formally possible to
talk of equilibrium below the glass transition, however all quantities
described below have been tested over different intervals at $T=0.25$
and found to be insensitive to the specific subset of events chosen
past event 400.  The exact value of the configurational energies after
thermalization with ART is significantly lower than that of models
relaxed with molecular dynamics.  Starting with a configuration
equilibrated at $T=1.5$ and slowly cooling down, (at $10^{-5}
\epsilon_{AA} /$ time unit), the inherent structures stabilize at an
energy per atom of about $-6.738$ at $T=1.00$, $-6.817$ at $T=0.50$
and $-6.858$ at $T=0.25$.  The corresponding values for the ART run
are $-6.810$, $-6.844$ and $-6.870$, respectively.  ART is not
expected to fully describe the liquid phase since it is event-based
and does not include entropic contributions.  This means that
distributions, such as those presented below, obtained with ART should
be sharper than those generated with MD in the liquid and supercooled
liquid.  ART and MD should meet in the solid phase where the dynamics
is almost exclusively activated.  This is the case in other
systems~\cite{asi} and also here, with only a small energy difference
between both techniques at $T=0.25$.  This difference can be explained,
in large part, by the much better sampling of the energy landscape achieved
by ART. 

\begin{figure}
     \vspace*{-0.5cm}
\epsfxsize=7.7cm \epsfbox{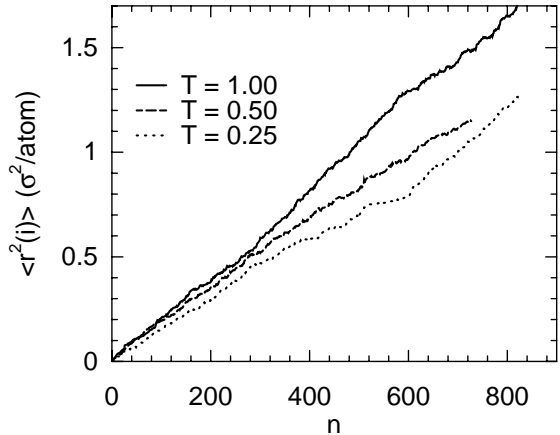}
\vspace{0.3cm}

\caption{ Mean-squared displacement per atom, from the initial
configuration, as a function of accepted event number above, around
and below the glass transition.  As mentioned in the text, the
diffusion due to atomic exchanges, which is significant at low
temperatures, is {\it not} included.  For comparison, a 2-million time
step MD simulation at $T=0.25$, with exchanges allowed, produces a
$\langle r^2\rangle = 0.035 \sigma^2$/atom.
\label{fig:diffusion}}
\end{figure}

At all temperatures, the sampling of the energy surface is
significant.  Figure \ref{fig:diffusion} shows the root-mean-square
displacement per atom as a function of accepted events.  Diffusion is
linear in this quantity, following an Einstein-like relation with a
``diffusion constant'' almost independent of the temperature.  In
terms of accepted events, i.e., without even including atomic
exchanges, the sampling of the phase space proceeds therefore at a
constant rate.  ART clearly overcomes the exponential slowing down of
the dynamics in these metallic glasses.

\begin{figure}

      \epsfxsize=8cm
      \epsfbox{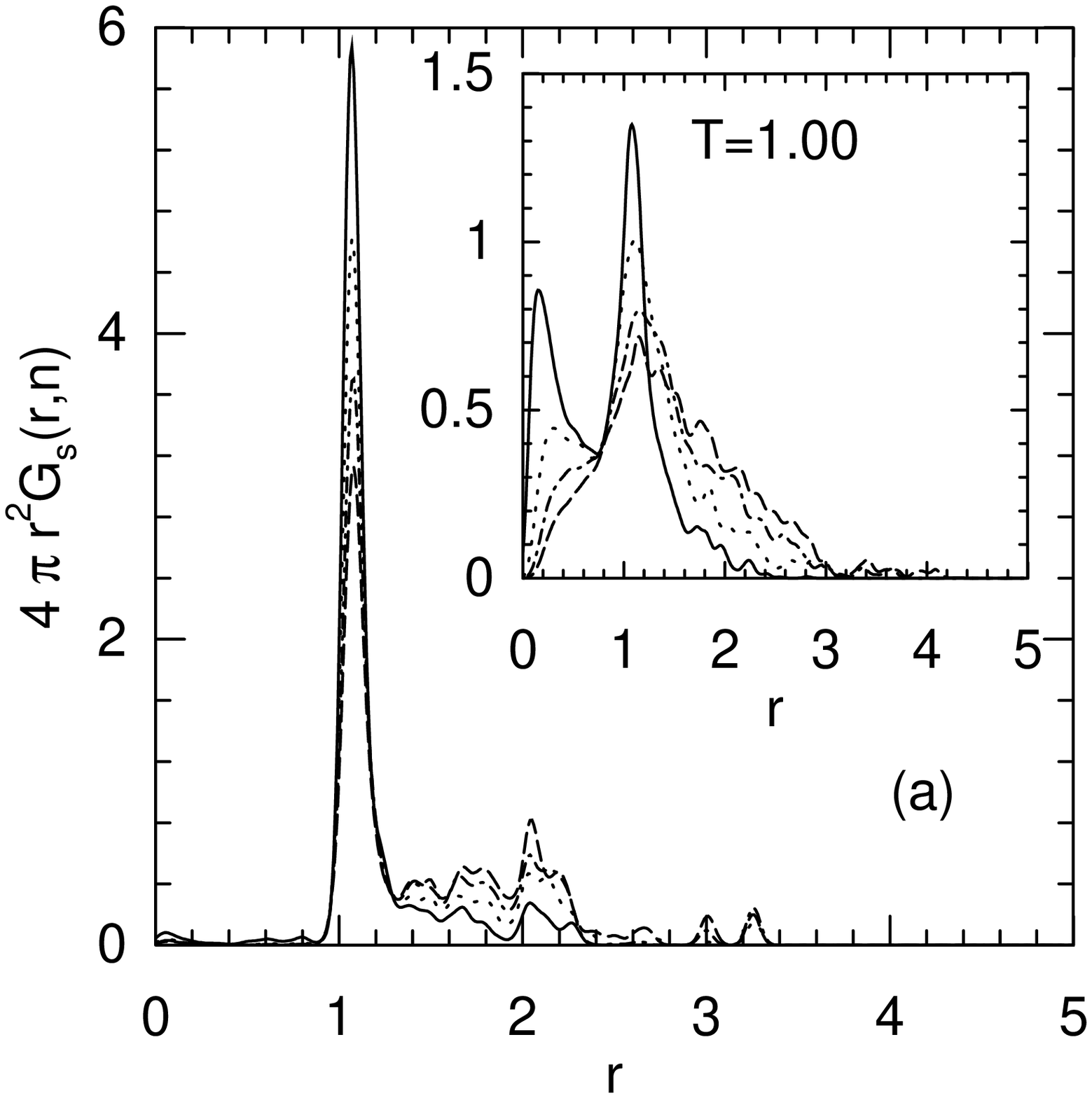}

      \epsfxsize=8cm
      \epsfbox{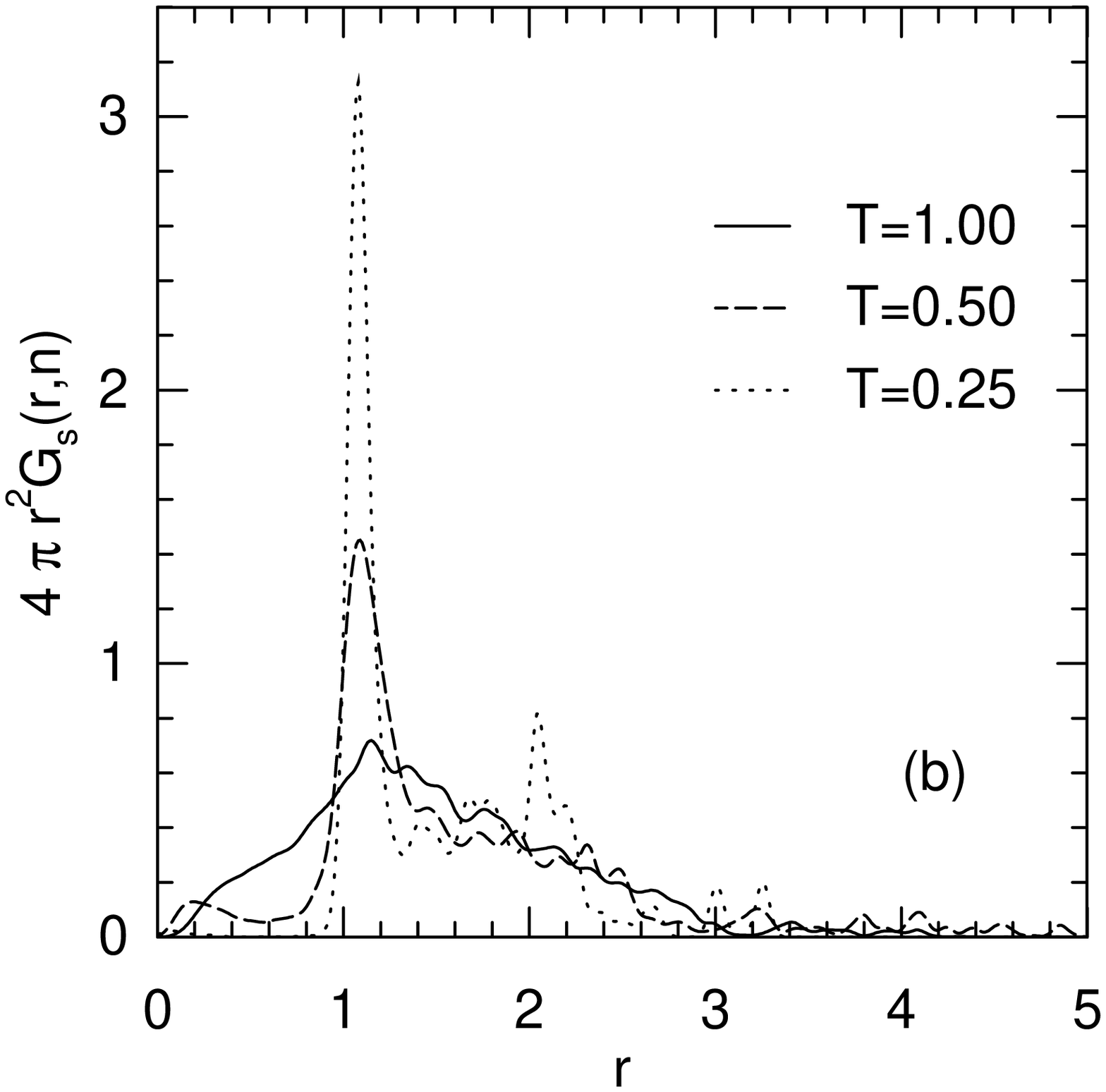}

      \vspace{0.3cm} 
      \caption{Self-part of the van Hove correlation
      function.  $n$ is taken as the difference between two
      configuration numbers.  (a) G$_{s}$ at T=0.25 for 4 different
      intervals: 20, 100, 200 and 300, starting with a solid line,
      dotted line, dash-dotted and dashed.  In inset, the same
      function is plotted for the run at $T=1.00$.  (b) Comparison of
      the G$_{s}$(r,t=300), at T=0.25, 0.50 and 1.00.  For both plots,
      the correlation function is measured in the second part of the
      run, i.e., after 300 accepted events.  The features of the
      correlation function are not sensitive to the subset of accepted
      events selected.}
      \label{fig:van}
\end{figure}

With the properties of the sampling established, we can study the the 
nature of the relaxation and diffusion mechanisms as a function of 
temperature.  The self part of the van Hove correlation function, 
\begin{equation}
     4 \pi r^2 G_s(r,n) = \frac{ 4 \pi r^2 }{N} \langle \sum_{i} 
\delta
\left( r - |{\bf r}_i(0)
      - {\bf r}_{i}(n) |\right) \rangle
\end{equation}
provides  the probability of finding an atom $r$ away from it initial 
position at event $n$. 

In Figure \ref{fig:van}(a), the correlation is plotted as a function 
of event number $n$ at two different temperatures.  The distributions at all 
temperatures are significantly sharper than those generally obtained 
in molecular dynamics \cite{lewis91,kob95a,donati98}.  Because ART 
concentrates exclusively on activated events, the thermal 
contribution to the peak around $r=0$ are eliminated.

\begin{figure}
      \epsfxsize=8cm
      \epsfbox{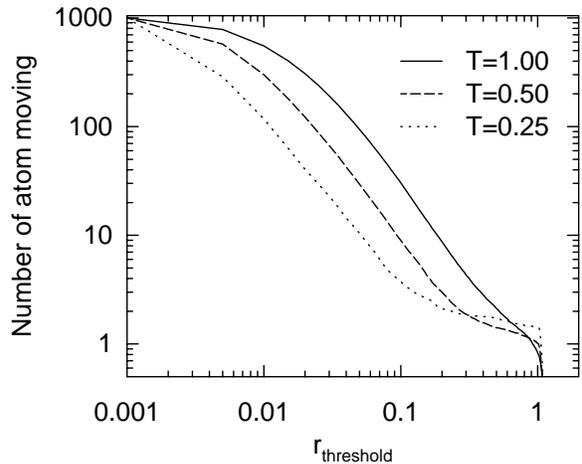}

      \vspace{1cm} 
      \caption{Log-log plot of the average number of atom per event
      moving by more than a given threshold.  At T=1.00, 30.4 atoms
      move by more $r=0.1\sigma$, while 8.9 do so at T=0.5 and only
      3.8 at T=0.25.  }
\label{fig:npart}
\end{figure}

At $T=1.00$ (inset), the distribution is initially bimodal with broad
peaks and evolves into a single wide peak, indicating relaxation
mechanisms with a range of of lengths.  The distribution at $T=0.25$,
on the other hand, is narrow and evolves by adding new peaks at
integer multiples of $r=1.05 \sigma$.  This behavior is underlined in
Fig.  \ref{fig:van}(b), which compares the long time distribution at
all temperatures: as the temperature lowers, the atomic displacements
become more and more discrete: the jumps take place on a disordered
but rigid lattice.  The $T=0.50$ model, well into the supercooled
region, is not yet as discrete: the distribution is broad and smooth
beyond the first peak.

In order to get a better understanding of the events, it is useful to 
plot the average number of atoms involved in an 
event.~\cite{mousseau00} Figure \ref{fig:npart} gives the full 
distribution as a function of a threshold displacement.  As discussed 
elsewhere,~\cite{mousseau00} the slope of the curve is universal.  
There is a strong temperature dependence on the pre-factor, however, 
which relates to the rigidity of the network sampled locally: the 
events become more and more collective as the temperature increases.  
At $r=0.1 \sigma$, events involve more than 2 times more atoms at 
$T=0.50$ and eight times more atoms at $T=1.00$ than at $T=0.25$.

over, there are essentially no atoms moving by $0.2 < r/\sigma <
1.0$ in the $T=0.25$-run, as indicated by a flat curve: atoms moving
have to jump by a nearest-neighbor distance, anything less is not
possible.  A similar phenomenon can be seen in the $T=0.50$ curve, but
for $0.6<r/\sigma<1.0$, and a slope still relatively high: as an atom
jumps by a nearest-neighbor distance, atoms in the local environment
relax by up to $r\/sigma=0.6$.  Although the degree of collectivity in 
the events varies strongly with temperature, the three curve come
together at $r=1.0 \sigma$.  At all temperatures, the activated
dynamics is therefore controlled by nearest-neighbor jumps. This
characteristic can also be seen in the self-part of the van Hove
correlation (Fig \ref{fig:van}):  for small intervals, $t=20$, the
$T=1.00$ distribution shows a strong peak at $r=1.05 \sigma$,
identical to that at $T=0.25$.

How do these results fit with those on heterogeneities in glasses 
obtained by MD?  The conclusions of these simulations 
can be summarized as follows: heterogeneities exist in glasses but 
they are not associated with a given scale; their spatial extent 
depends directly on the time scale selected, leading to homogeneities 
in the long run.  \cite{yamamoto98,donati99} Similar results are 
found here.  Comparing the first and the last half of the run at 
$T=0.25$, for example, we find that 226 atoms move by more than 
$r=0.2$ in the first half of the last 400 events, and 212 in the 
second, with 75 atoms belonging to the two sets, in agreement with MD 
work.

Where ART and MD differ, however, is in the definition of 
``collective'' event.\cite{donati98,donati99} Certain groups have 
concluded that the dynamics becomes more and more collective as the 
temperature decreases,\cite{natelson98,donati99} in apparent 
contraction with our results.  However, the adjective collective is 
applied there for events taking place over an extended time period 
and pertains more to time correlation between events than to the 
nature of each activated jump.  As such, ART and MD conclusions are 
not contradictory.  Further analysis is currently underway to see 
whether ART produces the same long time correlations.~\cite{inprep}

In summary, we have performed a simulation of a binary Lennard-Jones 
glass above and below the glass transition temperature.  We first 
show that ART can explore the phase space below $T_g$ without 
encountering exponential slowing down.  We also find that there is a 
qualitative change in the nature of the mechanisms as one goes from 
above to below T$_g$.  Although all events seem to be centered around 
a few atoms jumping over a lattice spacing, the total number of atoms 
involved in an event decreases by an order of magnitude as the 
temperature goes down.  These results support the interpretation of 
recent NMR experiments.~\cite{tang99} Although these results are 
likely to be valid for metallic glasses in general, we can expect 
different behavior from less dense glasses such as network glasses.

{\it Acknowledgments.} I would like to acknowledge numerous
discussions with G. T. Barkema, S. de Leeuw, Y. Limoge, Y. Wu, as well
as D. Drabold.  This work is supported in part by the NSF, under grant
DMR-9805848.  Part of these simulations where run on the computers of
HP$\alpha$C at TU Delft.

\bibliographystyle{prsty}

\begin{thebibliography}{99}
\bibitem{asi} In amorphous silicon, the energy surface sampled by ART
is also slightly below that reached by MD. Structural analysis reveals
however that the models created by ART are of much better quality,
emphasizing again that the energy difference for the solid phase is
essentially due to a better sampling and relaxation.  (Ref. 
\protect\cite{mousseau00}.)
\end{thebibliography}

\end{document}